\documentclass[11pt,a4paper]{article}
\usepackage[margin=25mm]{geometry}
\usepackage{fontspec}
\usepackage{fancyvrb}
\usepackage{tikz}
\usetikzlibrary{arrows.meta,positioning}
\usepackage{array}
\usepackage{longtable}
\usepackage{enumitem}
\setlist{itemsep=1pt, parsep=0pt, topsep=3pt}
\usepackage{cite}
\usepackage{url}
\usepackage[hidelinks]{hyperref}
\newcommand{\code}[1]{\texttt{#1}}

\title{CoordX Types: A Trustless, Swappable Type System\\for Shared Records}
\author{David Hyland-Wood, Rob Ewin and Peter Robinson\\CoordX}
\date{White paper, 9 November 2026}
\begin{document}
\maketitle

\begin{abstract}
The CoordX Protocol carries signed and encrypted entries between parties. It
proves who wrote each entry and who had permission to write it. It does not
say what an entry means. That task belongs above the protocol layer.

This paper describes the CoordX type system, the layer that gives entries a
meaning. A \textbf{type} is a short document that says how to read one part
of a shared record: which entries count, what each position means, what a
party may do there, and what a party must do next. The type system has two
key properties. It is \textbf{trustless}: a type is an ordinary signed
object fixed by its digest, and every party evaluates it for itself over
entries that the CoordX Protocol has already verified. No party has to trust
a server, a registry or another party's software. It is
\textbf{swappable}: a type names the interpreter it is written for, so many
interpreters, of different capabilities and designs, can be used side by
side in one record.

We describe how to write type schemas, and we define the expressive power of
the schema language. Payloads form a regular language. Entry structure forms
a regular tree language. Every rule and every computed value is a total
function in the primitive recursive class, so evaluation always halts.
We compare this with KQML, FIPA-ACL and CBCL, and we relate it to
language-theoretic security.
\end{abstract}


\smallskip\noindent\textbf{ACM CCS (1998):} C.2.4; D.3.1; F.4.3; I.2.11

\section{Introduction}\label{sec:intro}

The CoordX Protocol white paper~\cite{coordx} describes a protocol for shared
records with no central server. Each record is an \textbf{object}: a causal
tree~\cite{causal} of signed, immutable entries. Each entry has one parent. Each party
checks every entry by itself. It verifies each entry's validity including access permissions.
However, the contents of entries are opaque to the protocol. It does not know if 
an entry represents a vote, a payment, a comment or a chess move.

A workspace needs more than that. Three companies that sign a contract must
agree on what ``signed'' means. Two teams that share a task list must agree
on when a task is done. A group of software agents must agree on whose turn
it is. With a central server, the server's code holds these rules. Without a
central server, the rules must travel with the record, and each party must
run them for itself.

A \textbf{type} holds these rules. It is a short document, written in a
small schema language, that says how to read one typed subtree of an object. Every
party that reads the subtree under the same type gets the same meaning from
the same entries.

This paper makes four contributions:

\begin{enumerate}
\item It shows why the type system is \textbf{trustless}
  (Section~\ref{sec:trust}).
\item It shows why the type system is \textbf{swappable}: the type system is
  separate from any one interpreter (Section~\ref{sec:swap}).
\item It describes how to \textbf{write type schemas}
  (Sections~\ref{sec:schema} to~\ref{sec:cookbook}).
\item It states the \textbf{expressive power} of the schema language, and
  compares it with other agent communication languages
  (Sections~\ref{sec:expr} and~\ref{sec:acl}).
\end{enumerate}

This paper describes the type system and its schema language. It does not
describe how an interpreter is built. The schema language described here is
the one accepted by \code{interp/4}, the first interpreter that CoordX
publishes.

\section{Design principles}\label{sec:design}

Six principles define the type system.

\paragraph{1. Types.} A type is the payload of an
entry. Anybody can write and publish one. Nothing gives a type special
standing. A type names its interpreter and its dependencies by protocol
refs. The name that a type gives itself, such
as \code{tracker/1}, is only for display purposes. 

\paragraph{2. Meaning is positional.} 
The meaning of a type entry depends on entries in the tree between it and the root entry.
Each entry has a parent entry and can have zero or more children entries. 
The location of an entry within a tree is its \textbf{position}.

\paragraph{3. A reading is a pure function of what the reader holds.} 
The interpretation of a tree of entries is deterministic. 
It does not depend on external factors such as the time.
Interpretation is stateless, keeping nothing between calls.
As such, all parties with the same entries interpret the entries in precisely the same way.

\paragraph{4. A judgment is a fact.} A \textbf{rule} decides if an entry is
admitted. A rule reads only the entry's causal past: the entry, its
ancestors, and the entries it cites, transitively. A rule never reads
children, siblings or grants, because those can arrive at any later time.

\paragraph{5. Conflicts are shown, not resolved.} Rules cannot see
siblings. So two parties can each act from the same position, each adding entries. 
Both
entries can be admitted. The rule does not choose a winner. It says what the
two entries mean together: for example ``contended'', ``disputed'' or
``spent twice''.

\paragraph{6. An action is a plan.} A party never asks a server to ``do
something''. The type turns an action into a \textbf{plan}: an ordered list
of the exact entries and grants to write. Before anything is posted, the
proposer reads the tree with the plan's entries added. If the type would
refuse a planned entry, the proposer sees that first.

\section{Trustless by construction}\label{sec:trust}

``Trustless'' has a narrow meaning here. A party never has to accept a
statement because of who made it. Each party can check each statement for
itself, from bytes it holds. The type system gets this property in two
layers.

\paragraph{The protocol layer.} The CoordX Protocol~\cite{coordx} gives the
type system its inputs. Each entry is signed. Each entry names the
permission that allowed it. Any party can check the signature and the
permission without help. Content is encrypted, so relays see only sealed
bytes. Two honest parties that check the same bytes with the same keys get
the same answer. So the entries that a type reads are already facts.

\paragraph{The type layer.} The type system keeps that property as it adds
meaning:

\begin{itemize}
\item \textbf{Types are fixed by digest.} A record cites its type by a ref
  to exact bytes. Nobody can change a type in place. A changed type is a
  different type with a different digest. So the rules cannot change under
  a party after it has joined a record.
\item \textbf{There is no registry.} No party owns the list of types or the
  list of interpreters. A type is found by its ref, like any other entry.
\item \textbf{Every party evaluates.} Each party runs the type on its own
  client, over the entries it holds. Nobody sends a party ``the result''.
  A party that doubts a result computes it again.
\item \textbf{Evaluation is deterministic.} The same type and the same
  entries always give the same reading. There is no clock, no hidden state
  and no unordered choice. Randomness that an action needs is fixed into
  the plan before it is posted, so every party can check it.
\item \textbf{Judgments do not depend on the future.} A rule reads only the
  causal past. So a verdict cannot be changed by an entry that arrives
  later, and a party cannot be refused because of something it could not
  see.
\item \textbf{Partial knowledge is honest.} A party that cannot open an
  entry gets \code{unknown}, not a guess. A reading never claims more than
  the reader can check.
\end{itemize}

\begin{center}
\begin{tabular}{p{55mm} p{90mm}}
\hline
\textbf{A party does not have to trust\dots} & \textbf{because\dots} \\
\hline
a server or a relay & there is none in the semantics; relays carry sealed
  bytes only. \\
the author of a type & the type is public, fixed by digest, and chosen by
  the parties that use it. \\
another party's software & each party evaluates the type itself. \\
another party's reading & a reading is reproducible from the entries. \\
the order of arrival & a verdict reads only the causal past. The tree structure does not 
  depend on arrival order. \\
\hline
\end{tabular}
\end{center}

What remains is trust in what is chosen, not in who runs it. A party must
choose a type whose rules it accepts, and an interpreter implementation that
it is willing to run. Both choices are local, and both are visible to every
other party.

\section{Swappable interpreters}\label{sec:swap}

\begin{figure}[h]
\centering
\begin{tikzpicture}[
  box/.style={draw, rounded corners=2pt, align=center, minimum width=34mm,
    minimum height=9mm, font=\small},
  arr/.style={-{Stealth[length=2.2mm]}},
  node distance=8mm and 14mm]
\node[box] (i) {\textbf{interpreter A}\\e.g.\ \code{interp/4}};
\node[box, right=of i] (t) {\textbf{type entry}\\\code{tracker/1}};
\node[box, right=of t] (d) {\textbf{dependency}\\a conversation type};
\node[box, above=of d] (j) {\textbf{interpreter B}\\another design};
\node[box, below=of t] (r) {\textbf{instance root}\\object read under\\\code{tracker/1}};
\node[box, below=of d] (n) {\textbf{nested root}\\subtree read under\\the dependency};
\draw[arr] (t) -- node[above, font=\scriptsize]{ref} (i);
\draw[arr] (t) -- node[above, font=\scriptsize]{ref} (d);
\draw[arr] (d) -- node[right, font=\scriptsize]{ref} (j);
\draw[arr] (r) -- node[right, font=\scriptsize]{read under} (t);
\draw[arr] (n) -- node[right, font=\scriptsize]{read under} (d);
\draw[arr] (n) -- node[below, font=\scriptsize]{beneath} (r);
\end{tikzpicture}
\caption{Interpreters, types and instances are all entries, linked by refs.
A nested subtree can be read under a type for a different interpreter.}
\label{fig:refs}
\end{figure}
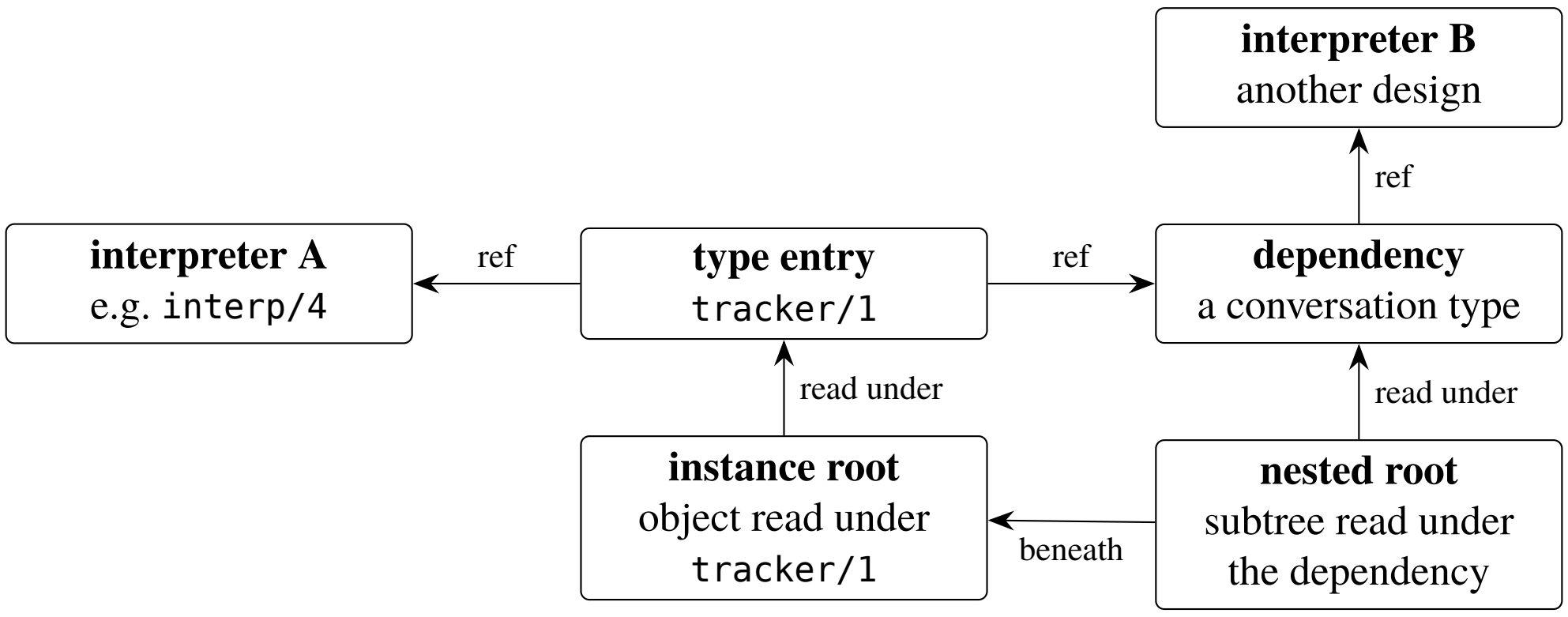

The type system and the interpreter are different things. The
\textbf{type system} is the model of this paper: positions, entry types,
rules, verdicts, views, obligations and plans. An \textbf{interpreter} is a
language, and the software that runs it. A type names
its interpreter by a ref, as Fig.~\ref{fig:refs} shows. The interpreter is
published as an ordinary entry, like the type.

The \textbf{runtime} is the software that hosts interpreters on a party's
client. Interpreters are plug-ins to the runtime. So:

\begin{itemize}
\item \textbf{Many interpreters can exist.} Anybody can publish one. Each is
  identified by its digest, in the same way as a type.
\item \textbf{Interpreters can differ.} One interpreter can be small and
  easy to audit. Another can have complex features such as more built-in functions.
\item \textbf{Interpreters can be mixed in one record.} A type can nest a
  subtree whose type is written for another interpreter. The runtime reads
  each subtree under its own type. An interpreter that a runtime does not
  have leaves that subtree unread. It does not make the enclosing type
  invalid.
\item \textbf{Implementations can be swapped.} Two implementations of one
  interpreter must give the same result. CoordX checks this with a
  conformance suite. A party can therefore choose the implementation it
  trusts, or write its own from a published interpreter definition.
\item \textbf{Old meaning is kept.} A type keeps its meaning under the
  interpreter it names. A new interpreter never changes an old type. A type
  moves to a new interpreter only when its author writes it again.
\end{itemize}

Swapping happens per type, not per reader. Two parties that read the same
subtree under the same type use the same interpreter definition, so they
get the same meaning. This is what lets the type system be swappable and
still trustless.

\section{Writing a type schema}\label{sec:schema}

The rest of this paper uses the schema language of \code{interp/4}. Authors
usually write a type in TOML~\cite{toml}. The authoring tool checks it and
writes the canonical JSON~\cite{rfc8259} that is published: members sorted
by key, no extra white space. Each expression is a string.

\subsection{The schema document}

\begin{center}
\begin{tabular}{>{\ttfamily}l p{105mm}}
\hline
\textrm{\textbf{Member}} & \textbf{Meaning} \\
\hline
schema & The name the type gives itself, for display. Required. \\
label & How the type is named to a person. \\
description & What the type is for. It has no semantics. \\
entries & The entry types (Section~\ref{sec:entries}). \\
state & The projections (Section~\ref{sec:proj}). \\
rules & The admission rules (Section~\ref{sec:rules}). \\
views & The views (Section~\ref{sec:views}). \\
exports & What an enclosing type can read (Section~\ref{sec:views}). \\
verbs & The actions (Sections~\ref{sec:verbs} and~\ref{sec:plans}). \\
create & How to start a new instance (Section~\ref{sec:create}). \\
\hline
\end{tabular}
\end{center}

The grammar is \textbf{closed}. A member that the grammar does not define
makes the schema invalid. A schema is checked in full when it is loaded,
before it reads any entry. A misspelt field, a rule that looks where it must
not, or a fold that could depend on order is refused at that point, with a
reason. So an error is found by the author, not by a reader.

\subsection{Values}

The language has \textbf{scalars}: integers (no other numbers), text,
booleans, parties, entries of this subtree, refs to entries of other
objects, and enumeration values. It has \textbf{collections}: lists (in
order), sets (no order) and maps. An \textbf{enumeration value} is a bare
upper-case name, with optional arguments: \code{DONE}, \code{WON(p)}. Entry
type names are enumeration values. A \textbf{party} is an identity. A type
can compare parties, put them in plans and records them in payloads. It can do
nothing else with them.

\paragraph{Two kinds of nothing.} The language separates two cases that most
languages join:

\begin{itemize}
\item \code{none} means \emph{absent}. A map key is missing, or an optional
  field is not given.
\item \code{unknown} means \emph{this reader cannot see it}. The entry is
  sealed (held but not readable), or content is not fetched yet.
\end{itemize}

\code{none} acts like ``false'' in \code{and} and \code{or}. \code{unknown}
spreads through every operation, even \code{==}. This is three-valued logic
in the style of Kleene~\cite{kleene}. A condition holds only when it is
exactly \code{true}. The function \code{known(x)} stops the spread. Use it
where the type means to go on with what this reader can see.

This split lets one type serve every reader. A reader that holds the whole
object and a reader that holds half of it run the same type. The second
reader gets \code{unknown} where it cannot see, not a wrong answer.

\subsection{Expressions}

Expressions use familiar forms: \code{x if c else y}, \code{and},
\code{or}, \code{not}, comparisons, \code{in}, integer arithmetic, field
access, index, slice, ranges \code{a..b} and calls. There are
comprehensions (\code{[e for x in S if c]}, and the set and map forms),
quantifiers (\code{any}, \code{all}, \code{count}, \code{sum}, \code{min},
\code{max}, \code{one}, \code{filter}) and a fold:
\code{fold(x in S: init -> step)}, with the accumulator named \code{acc}.

The set of functions is small and fixed: \code{known}, \code{len},
\code{keys}, \code{values}, \code{range}, \code{sorted}, \code{tip},
\code{deref}, \code{str}, \code{int}, and two sources of randomness that
only actions may use. There are no user-defined functions and no recursion.

An entry has fields: \code{id}, \code{parent}, \code{root}, \code{depth},
\code{path}, \code{children}, \code{descendants}, \code{author},
\code{type}, \code{grants}, \code{sealed}, \code{hypothetical}, its payload
fields and its projections. Which of these a part of the type may read is
limited (Section~\ref{sec:rules}).

\section{Entries and payloads}\label{sec:entries}

An entry type has a name, a list of the \textbf{parent} types it may sit
under, and a \textbf{payload}: named fields, each with a payload type. The
type \code{ROOT} names the subtree root.

\begin{center}
\begin{tabular}{>{\ttfamily}l p{100mm}}
\hline
\textrm{\textbf{Payload type}} & \textbf{Value} \\
\hline
int, string, bool & An integer, a text, a boolean. \\
Member & A party. \\
ref & An entry, of this object or another one. \\
{[T]}, \{T\} & A list or a set of \code{T}. \\
T? & An optional field. \\
\hline
\end{tabular}
\end{center}

A \code{ref} that names an entry of the same object reads as that entry.
That entry becomes part of the citing entry's causal past. This is how a
rule can reach beyond the path from the root.

An entry type can also declare a \code{label}, a one-line \code{summary}
(Section~\ref{sec:create}), and two flags:

\begin{itemize}
\item \code{private}: the entry is posted, and what is posted beneath it is
  readable only by the parties then granted there. The entry itself stays
  readable to everyone who reads its parent. This is how a type seals
  something now and reveals it later.
\item \code{subtree}: an entry of this type is the root of a nested subtree,
  read under a dependency type (Section~\ref{sec:nest}).
\end{itemize}

\section{Projections}\label{sec:proj}

A \textbf{projection} is a named value computed at a position. It is read
like a payload field: \code{self.balance} at an entry, \code{self.root.counts}
over the whole subtree. A projection gives exactly one of three forms:

\begin{itemize}
\item \code{value}: one expression.
\item \code{match}: the first arm whose \code{when} holds, with an optional
  \code{else}.
\item \code{fold}: a start value, then a step for each element of a
  collection.
\end{itemize}

A projection can be limited to some entry types with \code{on}.
Projections may read each other, but not in a cycle.

\paragraph{Folds and order.} Entry ids carry no meaning, so a result must
not depend on the order of a set. A fold over something that may be a set
must combine with \code{+} or \code{*} from an integer start, or with
\code{and} or \code{or} from a boolean start. Any other fold must be over a
list: a path, a range, a sorted collection or a list field. The loader
refuses any other fold.

\section{Rules and verdicts}\label{sec:rules}

A \textbf{rule} has an id, the entry types it applies to, and a
\code{require} expression. Each entry gets exactly one \textbf{verdict}:

\begin{center}
\begin{tabular}{>{\bfseries}l p{115mm}}
\hline
Verdict & Meaning \\
\hline
admitted & The entry counts. Its parent is admitted and every rule holds. \\
refused & The entry does not count. Its payload or type is wrong, it sits
  under the wrong parent, its parent is refused, or a rule fails. The
  verdict names the rule. \\
unverified & No rule fails, but at least one rule reached content that
  this reader cannot open. \\
waiting & Something the entry needs has not arrived yet. \\
sealed & The reader cannot open the entry. \\
\hline
\end{tabular}
\end{center}

A verdict is never firmer than its parent's. Verdicts only move forward. A
waiting entry is judged when what it waits for arrives. An unverified entry
becomes admitted or refused when its past becomes readable.

A refused entry stays in the tree, because the protocol admitted it. The
type leaves it out of \code{children}, \code{descendants}, views and
exports. Note the two layers. The protocol says an entry is valid. The type
says it does not count.

\subsection{What each part can see}

The most important rule for an author is that each part of a type sees only
part of the tree. The loader checks these limits.

\begin{center}
\begin{tabular}{p{40mm} p{55mm} p{50mm}}
\hline
\textbf{Part} & \textbf{Sees} & \textbf{Never sees} \\
\hline
a rule & the entry, its ancestors, and what it cites, transitively
  (its causal past) & children, descendants, siblings; grants; the reader \\
who may act (\code{for}), who must act (\code{of}) & the causal past of the
  position, and the reader & children, descendants, siblings; grants \\
a view; an obligation's \code{when}; an input's default and choices & the
  path to the position and everything beneath it & siblings of the position
  or of its ancestors \\
\hline
\end{tabular}
\end{center}

Three results follow. First, the root sees everything. Anything that must
take the whole object into account belongs at the root. Second, a lower
position is blind to its siblings. Third, a rule sees only the past. So a
rule cannot stop two parties from acting at the same position. It can only
check each entry against what came before it.

Authorship is a rule like any other: \code{self.author ==
self.parent.author}. There is no built-in owner.

\section{Views and exports}\label{sec:views}

A \textbf{view} belongs to a position. It has the entry types it appears
\code{at}, an optional \code{for} condition, a \code{show} expression and a
label. Views appear only at admitted or unverified entries.

\textbf{Values describe themselves.} Each value that a view gives has a
fixed display form: text, number, boolean, party, entry, none, unknown,
record (a map with text keys), or list. So a generic client can show any
type without code written for that type. Layout hints are not part of the
language.

An \textbf{export} is a named expression evaluated at the subtree root.
Exports are what an enclosing type can read of a nested subtree.

\section{Verbs: moves and obligations}\label{sec:verbs}

A \textbf{verb} is an action that a type offers at a position. Each verb
has a kind:

\begin{itemize}
\item A \textbf{move} is the reader's choice.
\item An \textbf{obligation} is a duty of the party that \code{of} names.
\end{itemize}

\paragraph{Offers.} A verb is offered to the reader at an entry of a type in
\code{at} when its \code{for} holds. An obligation is offered when
\code{of} is the reader. Because \code{for} and \code{of} read only the
causal past, who may act at a position is settled by its past.

An offer is presentation, not permission. A party may propose any declared
verb at any position. The rules judge the result. The only place that can
refuse an entry is a rule, and every reader runs the same rules.

\paragraph{Outstanding obligations.} An obligation is outstanding at an
entry when \code{of} gives a party and \code{when} holds. Unlike a rule,
\code{when} reads what is beneath the entry. So an obligation is discharged
by what is posted beneath it. Where one option excludes another, state one
obligation with a choice, not one obligation for each option.

\paragraph{Inputs.} A verb has an ordered list of named inputs. Each has a
type: \code{text}, \code{number}, \code{boolean}, \code{party},
\code{entry}, \code{options} (with \code{choices}) or \code{list}. A
\code{default} and \code{choices} are evaluated like a view at the position.

\section{Plans}\label{sec:plans}

A verb's body says what the action writes. Proposing the verb gives a
\textbf{plan}, or \textbf{not allowed} with a reason. A plan is data:
ordered steps, each naming exactly what it writes.

\begin{center}
\begin{tabular}{>{\ttfamily}l p{115mm}}
\hline
\textrm{\textbf{Step part}} & \textbf{What it does} \\
\hline
let & Binds names, in order. Randomness lives here. \\
create & Writes the root of a nested subtree under a dependency. \\
delegate & Proposes an action of an existing nested subtree, through that
  subtree's own type. \\
post, under, with & Writes one entry of a type, at a place, with payload
  fields. \\
each & Writes one entry for each element of a list. \\
grant & Gives parties kinds of permission at entries. \\
then & The next step. \\
\hline
\end{tabular}
\end{center}

Inside a step, \code{self} is the position, \code{reader} is the proposer
and \code{args} are the arguments. \code{\$entry} is what the step posted.
\code{\$entries}, \code{\$item} and \code{\$index} belong to \code{each}.

\paragraph{Placement.} A plan writes only at three places: at the position;
at an entry that the plan creates; or at an existing entry beneath the
position that the proposer names as an argument. A plan never picks among
existing entries by itself. Where to act is the proposer's decision. An
action that must also write somewhere unrelated is a separate action there.

\paragraph{Rehearsal.} Before a plan is posted, the proposer reads the tree
with the plan's entries added as hypothetical entries. A planned entry that
this reading refuses would be refused for real.

\paragraph{No atomicity.} A plan is several writes. Other parties receive
its entries in any order, and some may not arrive at all. A type must read
soundly for any subset of a plan's entries.

\section{Composition}\label{sec:nest}

A type can build on other types in two ways.

\paragraph{Nested subtrees.} An entry type with \code{subtree} cites a
dependency. An entry of that type is both an entry of this type and the
\code{ROOT} of a subtree read under the dependency. For example, a
\code{tracker/1} item carries a discussion read under a conversation type.
The enclosing type sees only where each nested root sits, which dependency it
cites, and its exports. A rule may read only those exports that cannot
change as entries arrive. If the enclosing type refuses the nested root, the
nested reading is discarded.

\paragraph{Other objects.} A \code{ref} can name an entry of another object.
\code{deref(r)} is that entry once the runtime has fetched it. Until then it
is \code{unknown}, and an entry whose rule needs it is \textbf{waiting}.

\section{Creating an instance and summaries}\label{sec:create}

The \code{create} member describes how to start a new object read under the
type. It has inputs, like a verb, and a \code{with} map that gives the
\code{ROOT} payload fields.

An entry type may declare a \textbf{summary}: an expression over the entry
alone, shown as one value. For example \code{["claimed by", self.author]}.
Summaries let a generic client list entries without reading the whole tree.

\section{A worked example: sign-off in order}\label{sec:example}

The type \code{signoff/1} is a document that needs approval from named
approvers, in order. The excerpt below shows the parts that matter.

\begin{Verbatim}[fontsize=\footnotesize, frame=single]
schema = "signoff/1"

[entries.ROOT]
payload = { approvers = "[Member]", title = "string" }
[entries.DRAFT]
parent = ["ROOT", "DRAFT"]
payload = { text = "string" }
[entries.DECISION]
parent = ["DRAFT", "DECISION"]
payload = { note = "string?", verdict = "string" }

[[rules]]
id = "APPROVERS"
on = "ROOT"
require = "len(self.approvers) > 0 and self.author not in self.approvers"
[[rules]]
id = "IN_TURN"
on = "DECISION"
require = "self.author == self.parent.next_approver"

[verbs.decide]
at = ["DRAFT", "DECISION"]
kind = "obligation"
of = "self.next_approver"
when = 'not any(c in self.children: c.type == DECISION)'
post = "DECISION"
with = { note = "args.note", verdict = "args.verdict" }
\end{Verbatim}

The type works like this:

\begin{enumerate}
\item The author creates the object. The rule \code{APPROVERS} refuses a
  root whose author is also an approver. Nobody approves their own work.
\item The author posts a \code{DRAFT} under the root. A revision is a
  \code{DRAFT} under a draft.
\item Each \code{DECISION} sits under the decision before it, so the
  decisions form a chain. The projection \code{next\_approver} reads that
  chain, which is the causal past of each position.
\item The rule \code{IN\_TURN} refuses a decision by anybody except the next
  approver. The rule reads only the parent, so its verdict is a fact.
\item The obligation \code{decide} names the next approver. It is
  outstanding while no decision sits beneath the position. When one arrives,
  the obligation goes away, and the next approver's obligation appears
  beneath it.
\item If two decisions sit under one position, no rule can see that, because
  they are siblings. The projection \code{status} at the draft sees both and
  shows the draft as \code{contended}.
\end{enumerate}

A \textbf{scenario} file plays the type with several parties. The
\code{signoff/1} scenario tries an approver who acts out of turn, and
expects \code{refused by IN\_TURN}. It then checks what each party sees and
owes at the end. The authoring tools start a type from a template, check it,
play scenarios, and explain why an expression reads as it does for one party
at one position.

\section{The cookbook}\label{sec:cookbook}

CoordX ships a cookbook of types. Each one shows patterns that most shared
records need.

\begin{center}
\begin{tabular}{>{\ttfamily}p{28mm} p{120mm}}
\hline
\textrm{\textbf{Type}} & \textbf{What it is} \\
\hline
poll/1 & The smallest type: people each add something once. \\
signoff/1 & A document approved in order (Section~\ref{sec:example}). \\
thread/3 & A conversation: each message under what it answers, citing what
  its author had seen. \\
thread/4 & \code{thread/3} with an end. After the end, nobody who has seen
  it can write. \\
tracker/1 & A work tracker: items and sub-items, claims, finishes, links,
  and a discussion nested under each item. \\
tokens/2 & A ledger of bearer notes: a handover sits under the holding it
  spends, so a note's history is its path. \\
cnet/1 & A call for proposals. Named contractors offer a price. A named
  timekeeper signs a close that cites the offers it received, and the lowest
  of those wins. \\
tictactoe/1, \newline rps/3, hangman/2 & Games. They show turns, sealed moves, and
  reveals. \\
\hline
\end{tabular}
\end{center}

\begin{center}
\begin{tabular}{p{50mm} p{95mm}}
\hline
\textbf{Pattern} & \textbf{Use it when} \\
\hline
Act at the root and name the target & An action must take the whole record
  into account but writes at one entry. \\
Carry a fact forward & A later position needs a fact from another branch.
  The entry records it, and a rule checks it against what it cites. \\
Cite, do not search & An entry must provably come after others. \\
Seal now, reveal later & Something must exist now but stay hidden: sealed
  bids, votes, an answer key. \\
A path is a sequence & Steps happen in order, each answering the last. \\
Two from one is read, not prevented & Two parties may act from one point.
  The reading says what that means. \\
Nobody approves their own & An approval must come from somebody else. \\
Obligations say what is next & A party must act for the record to go on. \\
\hline
\end{tabular}
\end{center}

\section{Expressive power}\label{sec:expr}

We state the power of the schema language at four levels. Each level is a
separate language with its own class. We give each result with a short
argument.

\paragraph{Schemas.} A schema is a canonical JSON document checked against
a closed grammar. JSON is a deterministic context-free language. Canonical
form gives each schema exactly one byte encoding, so two parsers cannot
disagree about which schema they hold.

\paragraph{Payloads: regular.} A payload type is built from five scalar
types by list, set and option, with a nesting depth that the schema fixes.
There is no recursive payload type. A language of documents of bounded
depth over a finite set of field names is regular. So each entry type's
payloads form a regular language, and a payload is recognised in full
before any rule reads it. A payload that does not match is refused.

\paragraph{Structure: a regular tree language.} Each entry type lists the
parent types it may sit under. This is a constraint between a client's label
and its parent's label only. The set of trees that it allows is a local tree
language, which is a subclass of the regular tree
languages~\cite{tata}.

\paragraph{Rules and projections: primitive recursive.} Every iteration in
the expression language runs over a finite collection that exists before the
iteration starts: a comprehension, a quantifier, a fold or a range. There
are no user-defined functions, no recursion, no unbounded loop and no cycle
between projections. So every expression halts on every input, and every
rule is a total, decidable predicate.

The class is exactly the primitive recursive functions. For the upper bound,
each construct is a bounded loop in the sense of the LOOP
language~\cite{meyer-ritchie}: its iteration count is fixed before the body
runs. For the lower bound, a fold over \code{range(0, n)} whose step is any
expression is a LOOP statement, and a fold whose start value is the outer
accumulator nests one loop inside another. LOOP programs compute exactly the
primitive recursive functions~\cite{meyer-ritchie}. So a type can compute
any primitive recursive function of its causal past, and nothing more.

\paragraph{Plans: bounded.} A plan is a finite chain of steps. The only
repetition is \code{each}, over a list fixed before it runs. So the writes
that one action can make are bounded by the type and its arguments.

\paragraph{What a type cannot express.} The limits are deliberate:

\begin{itemize}
\item \textbf{General computation.} No type is Turing complete. A type
  cannot loop until a condition holds.
\item \textbf{A global order.} A type cannot say which of two siblings came
  first. It can only show that there are two. An order must be built from
  citations and paths.
\item \textbf{Prevention of conflicts.} A rule sees only the past, so two
  parties that act at one position can both be admitted. The type must read
  the conflict and say what it means.
\item \textbf{Hidden state or time.} A type cannot read a clock, a network
  or anything outside the entries that a reader holds.
\end{itemize}

\paragraph{A note on cost.} Primitive recursive does not mean cheap. A
primitive recursive function can grow very fast. The guarantee is that
evaluation halts and is the same for every reader. The cost of a reading is
bounded by a function fixed by the type, not chosen by the party that sends
an entry. A party chooses a type before it accepts entries under it, so it
can refuse a type whose cost it does not accept.

\section{Comparison with agent communication languages}\label{sec:acl}

\subsection{Language-theoretic security}

Language-theoretic security (LangSec)~\cite{sassaman13} treats every input
as a language. Its main claims are three. First, an input must be
recognised in full before it is processed. Second, the input language must
be no more powerful than the task needs, because the power of a recogniser
above the deterministic context-free class makes parser equivalence
undecidable. Third, an input handler with more power than intended becomes a
``weird machine''~\cite{bratus11} that an attacker can program.

CBCL~\cite{cbcl} applies LangSec to agent communication. It limits every
message, including the definitions of new dialects, to the deterministic
context-free class. Three invariants, checked in Lean, bound expansion, apply
resource limits and keep the core vocabulary. CBCL's literature review
classifies earlier agent communication languages by the complexity of their
interpretation pipeline. We extend that classification to CoordX types.

\subsection{KQML and FIPA-ACL}

KQML~\cite{kqml} introduced speech-act \textbf{performatives}, such as
\code{tell} and \code{ask-if}, as the envelope of a message. It does not fix
the \textbf{content language}. Its authors state that it is too early to
standardise on one. They name KIF as a common content language, and KIF is
``based on first order logic'' with extensions~\cite{kqml}.

FIPA-ACL defines 22 communicative acts~\cite{fipa-cal}. A message names its
content language in a \code{language} parameter, and FIPA does not fix one
language~\cite{fipa-acl}. The semantics of each act is given in FIPA SL. SL
is ``a logic of mental attitudes and actions, formalised in a first order
modal language with identity''. Its primitive attitudes are belief,
uncertainty and choice~\cite{fipa-cal}.

KIF and SL both contain first-order logic, so validity in either is
undecidable. CBCL notes that both ACLs use an S-expression envelope, which
is deterministic context-free, but that neither constrains the content
language. It classes both as context-free, with partial verification and
partial extensibility~\cite{cbcl}.

Three further differences matter here:

\begin{itemize}
\item \textbf{Semantics that cannot be checked.} The FIPA-ACL semantics
  is stated in terms of the sender's belief, uncertainty and choice.
  Another agent cannot check a mental state. Singh~\cite{singh98} identified
  this reliance on ``unverifiable mental states rather than observable
  social commitments'', as CBCL puts it~\cite{cbcl}.
  CoordX types are public in that sense. A verdict, a view and an
  obligation are computed from signed entries that every party can check.
  An obligation is close to Singh's social commitment, but it is computed,
  not asserted.
\item \textbf{Trusted infrastructure.} KQML relies on facilitators: agents
  that keep a registry of service names, forward messages and route them by
  content~\cite{kqml}. Neither the KQML paper nor the FIPA message structure
  specification~\cite{fipa-acl} defines how a message is authenticated.
  CoordX has no facilitator. Every entry is signed and every party checks
  it.
\item \textbf{Messages, not records.} An ACL message is sent and then gone.
  A FIPA message can name an interaction protocol and a conversation, but
  the protocol is defined in another specification~\cite{fipa-acl}. A
  CoordX type holds the protocol, the record and the obligations together,
  in one object that every party holds.
\end{itemize}

\subsection{LLM-X}

LLM-X~\cite{llmx} is an exchange for messages between personal LLM agents.
Agents connect through gateways. A gateway checks a JWT bearer token,
validates the message against a JSON Schema, and limits its rate. The
message types follow the Contract Net: CFP, Offer, Accept, Reject and Ack.
A Policy Engine decides when the initiator selects among offers: at the
first valid offer, after a bounded wait, or after all offers arrive. A
Control Plane handles registration, consent and capabilities. The paper
measures latency and message volume with 5 to 12 agents, in runs of up to
12 hours. Its agents are Python wrappers. The authors list the integration
of real LLMs as future work~\cite{llmx}.

LLM-X and CoordX types both give agents typed messages for coordination.
They differ in where the checks run:

\begin{itemize}
\item \textbf{Who checks.} In LLM-X the gateway validates the schema and 
he Policy Engine applies the decision rules. Receivers must trust both. The
CoordX protocol and types are trustless. In CoordX every party processes
types themselves.
\item \textbf{Who wrote it.} An LLM-X gateway checks a bearer token when a
  message arrives. The paper does not describe a signature on the message
  itself~\cite{llmx}. A CoordX entry is signed, and every party checks the
  signature from the bytes that it holds.
\item \textbf{Where the rule lives.} An LLM-X policy is a setting of the
  Policy Engine. A CoordX rule is part of the type, which is fixed by its
  digest and read by every party.
\item \textbf{Time.} An LLM-X CFP carries a deadline, which the exchange
  applies. A CoordX type cannot read a clock (Section~\ref{sec:expr}). A
  type that needs a deadline names a party that posts a signed entry to
  close the round. The deadline is then a claim of that party, and every
  party can see who made it.
\end{itemize}

LLM-X reports performance under load. This paper makes no claim about
throughput or latency.

\subsection{Extension}

CBCL extends itself by dialects. A dialect is a message, and the
invariants keep it safe. KQML lets communities agree on new performatives~\cite{cbcl}.
FIPA-ACL allows user-defined parameters with the prefix \code{X-}, but FIPA
does not define their semantics~\cite{fipa-acl}. Neither checks an
extension.

CoordX extends at two levels. A new type is an ordinary object, checked in
full when it is loaded. A new interpreter is also an ordinary object, and
types choose it by ref. Neither can change the meaning of an existing type,
because each type is fixed by its digest. This plays the part of CBCL's
invariant that keeps the core vocabulary.

\subsection{Summary}

\begin{center}
\small
\begin{tabular}{>{\raggedright}p{20mm} >{\raggedright}p{24mm} >{\raggedright}p{25mm} >{\raggedright}p{22mm} >{\raggedright}p{22mm} p{20mm}}
\hline
\textbf{Language} & \textbf{Input class} & \textbf{Semantics} & \textbf{Checkable by receiver} & \textbf{Extension} & \textbf{Trust} \\
\hline
KQML & CFL envelope; content open & speech acts & partial & by agreement & facilitators \\
FIPA-ACL & CFL envelope; content open (e.g.\ SL) & mental states & partial & \code{X-} parameters & not specified \\
MCP & unbounded (JSON-RPC) & tool calls & no & yes & server \\
CBCL & DCFL, including dialects & message forms & yes & dialects, bounded & not addressed \\
LLM-X & JSON, checked by JSON Schema at a gateway & Contract Net acts; engine policy & schema only & new schemas & gateway, Policy Engine \\
CoordX types & regular payloads; regular tree structure; primitive recursive rules & public, computed from signed entries & yes, by every party & new types and new interpreters, by digest & none beyond the protocol \\
\hline
\end{tabular}
\end{center}

The MCP row follows CBCL's own table~\cite{cbcl}. ``Trust'' is what a party
must rely on beyond its own checks.

The comparison has one caveat. CBCL limits the language at the parser.
CoordX types limit it at the parser and also limit the computation that
gives an entry its meaning. That computation is above the context-free
class, but it is total. So CoordX accepts more power than CBCL in the rules,
in exchange for meaning that every party can compute and check. It accepts
much less power than an ACL whose content is open, or a tool protocol whose
handlers can run any code.

\section{Limits}\label{sec:limits}

\begin{itemize}
\item \textbf{Plans are not atomic.} A reader can hold part of a plan. A type
  must read soundly for any subset.
\item \textbf{Readings differ by reader.} That is correct, not a fault. A
  reading is never the state of the object, and a client must not present it
  as one.
\item \textbf{A sealed entry hides its type.} A reader sees only the set of
  types its parent admits there. Rules that reach it give unverified.
\item \textbf{Silent none.} In \code{interp/4}, a view or condition that
  reaches outside its position's sight gets \code{none}, not
  \code{unknown}. A later interpreter may change this.
\item \textbf{Silent failure of obligations.} In \code{interp/4}, an
  obligation whose \code{of} or \code{when} fails to evaluate is left out.
  The record then shows nothing owed.
\item \textbf{Mixed interpreters share little.} A nested subtree under
  another interpreter exports nothing that \code{interp/4} can read.
\end{itemize}

Interpreter-specific limits like these are exactly what swappability is
for. A later interpreter can remove them without changing any existing
type.

\section{Status}\label{sec:status}

\code{interp/4} is frozen. It has a reference implementation that the
runtime loads as a plug-in, a conformance suite, authoring tools, and the
cookbook. Proposals for a later interpreter are under review. They do not
change \code{interp/4} or any type written for it.

\section{Conclusion}\label{sec:conclusion}

The CoordX Protocol proves who wrote an entry and who could. The type system
says what the entry means. It is trustless: a type is an ordinary object
fixed by its digest, every party evaluates it, and every judgment reads only
a past that every party can check. It is swappable: a type names its
interpreter, so interpreters of different designs can coexist and improve
without breaking old types. Its schema language is deliberately small.
Payloads are regular, structure is a regular tree language, and every rule
is primitive recursive and so always halts. That places CoordX types with
CBCL on the safe side of the LangSec line, and well inside the power of
KQML and FIPA-ACL, while it adds what neither of those has: meaning that
every party computes for itself.

\section{Acknowledgements}\label{sec:acknowledgements}

Thanks to Peter Robinson for his reviews and contributions to the type
system description.

\appendix

\section{Terms}\label{app:terms}

\begin{description}[style=nextline, itemsep=2pt]
\item[causal past] An entry, its ancestors, and what it cites by ref,
  transitively. The only part of the tree that a rule reads.
\item[dependency] A type that another type cites by ref, to type a nested
  subtree.
\item[export] A named value of a subtree, evaluated at its root, that an
  enclosing type can read.
\item[interpreter] A language for types, and the software that runs it,
  published as an entry and named by a type. \code{interp/4} is one.
\item[move] A verb that the reader may choose to do.
\item[obligation] A verb that a named party must do. It is outstanding until
  something beneath the position discharges it.
\item[plan] The exact ordered writes that carry out an action.
\item[position] An entry, seen as the place where a meaning, a view or an
  action belongs.
\item[primitive recursive] The class of functions computable with loops
  whose number of runs is fixed before they start.
\item[projection] A named value computed at a position.
\item[reading] What one reader's client knows of a subtree: verdicts, views,
  exports and obligations.
\item[rehearsal] Reading the tree with a plan's entries added as
  hypothetical, before posting.
\item[rule] An expression that must hold for an entry to be admitted.
\item[runtime] The software on a party's client that hosts interpreters.
\item[sealed] An entry that the reader holds but cannot open.
\item[swappable] Able to use many interpreters, chosen per type.
\item[trustless] Able to check every result from bytes one holds, without
  relying on who made it.
\item[type] A schema that says how to read one typed subtree.
\item[type system] The model of this paper: positions, entry types, rules,
  verdicts, views, obligations and plans.
\item[unknown] A value that this reader cannot see. Different from
  \code{none}, which is absent.
\item[verdict] The judgment of one entry: admitted, refused, unverified,
  waiting or sealed.
\item[view] A named, self-describing value shown at a position.
\end{description}

\end{document}